\documentclass[aps,prl,showpacs,amsmath,amssymb,amsfonts,superscriptaddress,twocolumn,lengthcheck]{revtex4-2}
\usepackage{ulem}
\usepackage{color}
\usepackage{ulem}
\usepackage{amsmath}
\usepackage{amssymb}
\usepackage{graphics}
\usepackage{graphicx}
\usepackage{dcolumn}
\usepackage[pdfpagemode=UseNone,pdfstartview=FitH,colorlinks=true,linkcolor=blue,urlcolor=blue,anchorcolor=blue,citecolor=blue]{hyperref}
\usepackage{booktabs}

\begin{document}
\title{Magnon squeezing in the quantum regime}

\author{Yuan-Chao Weng}
\thanks{These authors contributed equally to this work.}
\affiliation{Zhejiang Key Laboratory of Micro-Nano Quantum Chips and Quantum Control, School of Physics, and State Key Laboratory for Extreme Photonics and Instrumentation, Zhejiang University, Hangzhou 310027, China}

\author{Da Xu}
\thanks{These authors contributed equally to this work.}
\thanks{Email:~daxu@zju.edu.cn}
\affiliation{Zhejiang Key Laboratory of Micro-Nano Quantum Chips and Quantum Control, School of Physics, and State Key Laboratory for Extreme Photonics and Instrumentation, Zhejiang University, Hangzhou 310027, China}

\author{Zhen Chen}
\thanks{These authors contributed equally to this work.}
\affiliation{Beijing Key Laboratory of Fault-Tolerant Quantum Computing, Beijing Academy of Quantum Information Sciences, Beijing 100193, China}

\author{Li-Zhou Tan}
\affiliation{Zhejiang Key Laboratory of Micro-Nano Quantum Chips and Quantum Control, School of Physics, and State Key Laboratory for Extreme Photonics and Instrumentation, Zhejiang University, Hangzhou 310027, China}

\author{Xu-Ke Gu}
\affiliation{Zhejiang Key Laboratory of Micro-Nano Quantum Chips and Quantum Control, School of Physics, and State Key Laboratory for Extreme Photonics and Instrumentation, Zhejiang University, Hangzhou 310027, China}

\author{Jie Li}
\affiliation{Zhejiang Key Laboratory of Micro-Nano Quantum Chips and Quantum Control, School of Physics, and State Key Laboratory for Extreme Photonics and Instrumentation, Zhejiang University, Hangzhou 310027, China}

\author{Hai-Feng Yu}
\thanks{Email:~hfyu@baqis.ac.cn}
\affiliation{Beijing Key Laboratory of Fault-Tolerant Quantum Computing, Beijing Academy of Quantum Information Sciences, Beijing 100193, China}

\author{Shi-Yao Zhu}
\affiliation{Zhejiang Key Laboratory of Micro-Nano Quantum Chips and Quantum Control, School of Physics, and State Key Laboratory for Extreme Photonics and Instrumentation, Zhejiang University, Hangzhou 310027, China}

\author{Xuedong Hu}
\affiliation{Department of Physics, University at Buffalo, SUNY, Buffalo, NY, 14260-1500, USA}

\author{Franco Nori}
\affiliation{Quantum Computing Center, RIKEN, Wakoshi, Saitama 351-0198, Japan}
\affiliation{Department of Physics, University of Michigan, Ann Arbor, Michigan 48109-1040, USA}

\author{J. Q. You}
\thanks{Email:~jqyou@zju.edu.cn}

\affiliation{Zhejiang Key Laboratory of Micro-Nano Quantum Chips and Quantum Control, School of Physics, and State Key Laboratory for Extreme Photonics and Instrumentation, Zhejiang University, Hangzhou 310027, China}

\begin{abstract} 
Squeezed states, crucial for quantum metrology and emerging quantum technologies, have been demonstrated in various platforms, but quantum squeezing of magnons in macroscopic spin systems remains elusive. Here we report the experimental observation of quantum-level magnon squeezing in a millimeter-scale yttrium iron garnet (YIG) sphere. By engineering a strong dispersive magnon-superconducting qubit coupling via a microwave cavity, we implement a significant self-Kerr nonlinearity to generate squeezed magnon states with their mean magnon number less than one. Harnessing a magnon-assisted Raman process, we perform Wigner tomography, revealing quadrature variances of $\sim\!0.8$ ($\sim\!1.0$~dB squeezing) relative to the vacuum. These results lay the groundwork for quantum nonlinear magnonics and promise potential applications in quantum metrology.
\end{abstract}


\maketitle

\section{Introduction}

Squeezed states -- quantum states with reduced fluctuations in one quadrature at the expense of amplified fluctuations in the conjugate counterpart -- serve as foundational resources for quantum-enhanced metrology and nonclassical state engineering~\cite{Walls-Nature-1983}. These states can offer enhanced measurement sensitivities beyond the standard quantum limit in, e.g., gravitational-wave detectors~\cite{Abadie-Np-2011} and axion dark matter searches~\cite{Backes-Nature-2021}, while also enabling advances in continuous-variable quantum computing~\cite{Braunstein-RMP-2005,Weedbrook-RMP-2012} and quantum gravity tests~\cite{Bose-RMP-2025}. 

Recently, millimeter-scale yttrium iron garnet (YIG) ferrimagnets have emerged as a compelling platform for quantum technology development~\cite{Flatte-PRL-2010,Huebl-PRL-2013,Nakamura-PRL-2014,Tang-PRL-2014,Tobar-PRApplied-2014,Hu-PRL-2015,Bauer-PRB-2015} due to the high coherence of their Kittel mode (a uniform spin-precession magnon mode)~\cite{Zhang-QI-2015}, establishing them as an ideal candidate for investigating the quantum-classical boundary, macroscopic quantum phenomena, and emerging quantum applications~\cite{Lachance-APE-2019,Yuan-PR-2022,Rameshti-PR-2022}. Following the groundbreaking demonstration of coherent magnon-superconducting qubit coupling~\cite{Tabuchi-Science-2015}, rapid progress has been achieved in quantum magnonics, including techniques for magnon detection~\cite{Lachance-Quirion-SA-2017, Lachance-quirion-Science-2020, Rani-arxiv}, deterministic preparation of classical magnon states such as the Glauber coherent state $|\alpha\rangle$~\cite{Tabuchi-Science-2015}, as well as elementary quantum states such as the single-magnon state $|1\rangle$, its coherent superposition with the vacuum state $|0\rangle$, and the magnon-qubit entangled state~\cite{Xu-PRL-2023,Xu-QST-2024}. More exotic states, such as quantum-level squeezed states of magnons, remain elusive due to the need for strong nonlinear interactions to generate observable effects in the quantum regime~\cite{Rameshti-PR-2022,Elyasi-PRB-2020,Kamra-APL-2020}.

In this work, we experimentally demonstrate the quantum-level magnon squeezing in a macroscopic, at 1~mm-diameter, YIG sphere. By integrating the YIG sphere with a superconducting qubit inside a three-dimensional (3D) microwave cavity, we establish a strong dispersive magnon-qubit coupling~\cite{Imamoglu-PRL-2009,Tabuchi-Science-2015}. Similar to a recent nonlinear mechanical resonator~\cite{Marti-NP-2024}, this coupling induces a significant self-Kerr nonlinearity in the magnon system, enabling the generation of squeezed magnon states in the quantum regime. To fully characterize the magnon squeezing, we leverage a novel magnon-assisted Raman process~\cite{Pechal-PRX-2014,supplemental} to achieve in situ tunable magnon-qubit state swapping, which in turn facilitates direct Wigner tomography of the squeezed states. We obtain a minimum quadrature variance of $\sim\!0.8$ relative to the vacuum (corresponding to $\sim\!1.0$ dB of squeezing), and a mean magnon population below one, marking an unambiguous demonstration of the magnon squeezing in the quantum regime. Moreover, by maintaining the self-Kerr nonlinearity, we actively preserve a fragile squeezed state for 400 ns, substantially outliving the magnon's intrinsic lifetime of $\sim\!145$~ns. These advances hold promise for expanding the scope of quantum metrology, with potential applications ranging from gravitational-wave detection~\cite{Abadie-Np-2011} to axion dark matter searches~\cite{Crescini-PRL-2020,Mitridate-PRD-2020,Backes-Nature-2021}. Moreover, our work lays the groundwork for exploring quantum nonlinear magnonics, offering a platform to investigate novel nonlinear phenomena of magnons in the quantum regime.

\begin{figure*}
	\centering
	\includegraphics[width=0.95\textwidth]{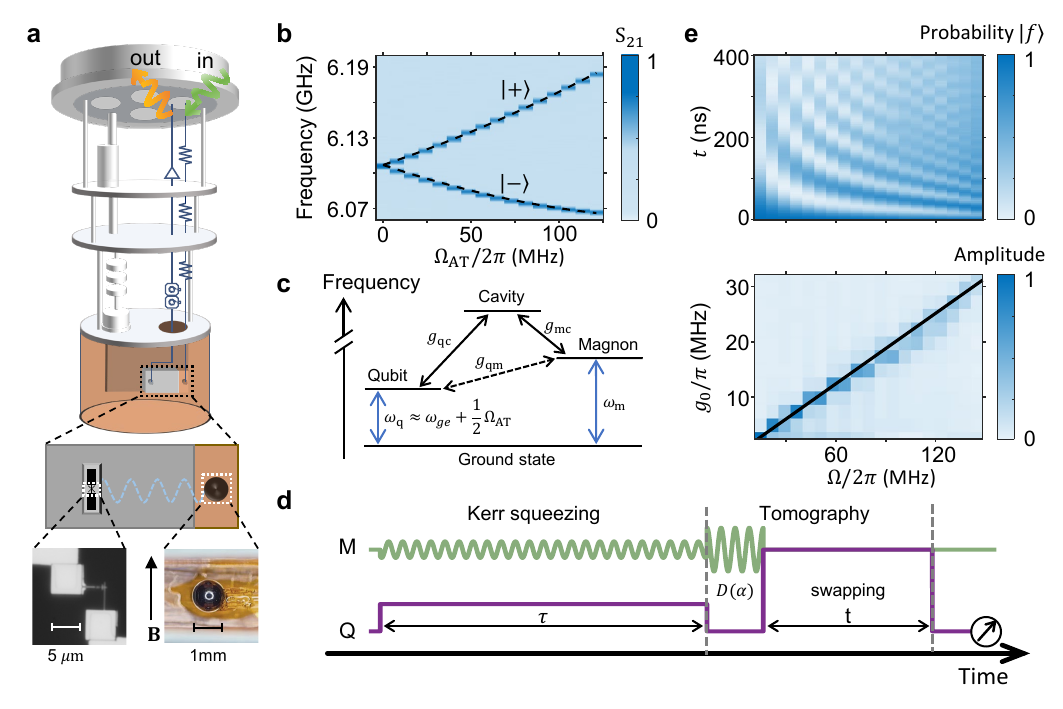}%
	\caption{\textbf{Experimental platform and qubit-magnon control scheme.}
		\textbf{a,}~Schematic of the hybrid quantum system, where a 1-mm-diameter YIG sphere and a transmon qubit are placed at the magnetic and electric field antinodes of the $\rm{TE}_{102}$ mode of a 3D rectangular microwave cavity, respectively. The system is operated in a dilution refrigerator at $\sim$10~mK. An external magnetic field $\mathbf{B}$ magnetizes the sphere, biasing the Kittel mode to $\omega_{\rm m}/2\pi = 6.231$~GHz.
		\textbf{b,}~Autler-Townes (AT) splitting of the qubit. The measured qubit spectrum is plotted as a function of the AT drive amplitude $\Omega_{\rm AT}$. The splitting forms two dressed states, $\vert + \rangle$ and $\vert - \rangle$, with the transitions from $\vert g \rangle$ to them shown as the upper and lower branches. Dashed lines are numerical fits.
		\textbf{c,}~Energy level diagram of the hybrid quantum system. The cavity mode is far-detuned from both the qubit and the magnon, mediating an effective qubit-magnon coupling $g_{\rm qm}$. A large detuning, $\Delta_{\rm qm} \approx 3.8 g_{\rm qm}$, ensures that the qubit-magnon system is in the dispersive regime and induces a strong self-Kerr nonlinearity on the magnons.
		\textbf{d,}~Pulse sequence for squeezing and tomography. The sequence is divided into three parts: (i) preparation of the squeezed state via Kerr nonlinearity, (ii) Wigner tomography of the magnon state, and (iii) final measurement of the qubit. Frequencies of the qubit (purple) and magnon (green) are indicated, along with the required microwave control pulses (wavy lines).
		\textbf{e,}~Tunable magnon-qubit state swapping. Top: A magnon-assisted Raman drive induces coherent Rabi oscillations between $\vert f,0 \rangle$ and $\vert g,1 \rangle$. The population in $\vert f,0 \rangle$ is plotted versus the interaction time. Bottom: The effective swap coupling strength $g_0$, extracted by FFT, is shown to be linearly tunable with the Raman drive amplitude $\Omega$. The solid line is a linear fit.}
	\label{fig1}
\end{figure*}

\begin{figure*}
	\includegraphics[width=0.98\textwidth]{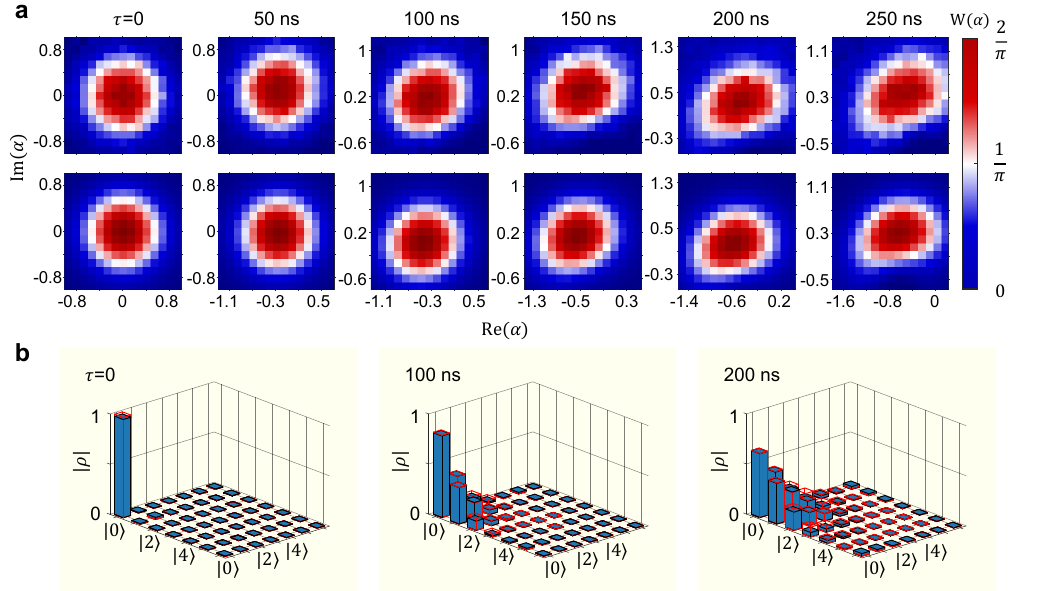}%
	\caption{\textbf{Wigner tomography and density matrices of the vacuum and squeezed states of magnons.} \textbf{a,}~Top row: experimentally measured Wigner function for the vacuum state ($\tau=0$) and for five squeezed states at $\tau=50$, 100, 150, 200, and 250~ns. Bottom row: the corresponding Wigner functions derived from the numerically simulated density matrices. Both experimental and numerical data clearly illustrate the transition from a vacuum state (circular distribution) to nonclassical squeezed states (quasi-elliptical distributions), with the strongest squeezing emerging at around $\tau=150$~ns, where the minimum variance $V_{\mathrm{min}}(\tilde{X})\approx0.799 \pm 0.068$ ($\sim\!1.0$~dB).
		\textbf{b,} Density matrices (blue bars) for the vacuum state at $\tau=0$ and two squeezed states at $\tau=100$ and 200~ns. The red bars represent the corresponding numerical simulations (see \cite{supplemental} for details). Each density matrix is truncated at the magnon Fock state $|n\rangle$ with $n=6$ for clarity.}
	\label{fig2}
\end{figure*}

\section{Results}

\subsection{Experimental setup and magnon squeezing protocol}

The intrinsic Kerr nonlinearity of the Kittel mode in the YIG sphere is negligibly small ($\sim 0.1$~nHz)~\cite{Zhang-SCPMA-2019}, rendering it insufficient for generating magnon squeezing in the quantum regime. To overcome this limitation, we engineer an effective Kerr nonlinearity by coupling the YIG sphere to a superconducting transmon qubit that acts as a strongly nonlinear element, with both embedded in a 3D microwave cavity~\cite{supplemental}, which mediates the magnon-qubit interaction (Fig.~1a). We achieve here a substantial enhancement of the magnon-qubit coupling strength, with a system cooperativity $C=\frac{4g_{\rm  qm}^2}{\gamma_{\rm q}\gamma_{\rm m}} \sim 1.0\times10^4$, representing an order-of-magnitude improvement over the previous implementations~\cite{Xu-PRL-2023,Xu-QST-2024}.  This strong coupling leads to a significantly increased Kerr nonlinearity to produce observable magnon squeezing in the quantum regime. Note that the spatially uniform Kittel mode is the only magnon mode strongly coupled to the cavity and hence the qubit~\cite{Tabuchi-Science-2015,Tabuchi-CRP-2016}, ensuring here the squeezing protocol targets the Kittel mode.

In our experimental setup, the Kittel mode frequency is fixed at $\omega_{\rm m}/2\pi=~6.231~\rm{GHz}$, while the transmon transition frequencies are $\omega_{ge}/2\pi=6.107~\rm{GHz}$ and $\omega_{ef}/2\pi=5.858~\rm{GHz}$. The qubit anharmonicity is given by $\eta/2\pi=(\omega_{ef}-\omega_{ge})/2\pi=-249~\rm{MHz}$, where the subscripts $g$, $e$ and $f$ refer to the ground, first-excited and second-excited states of the transmon, respectively. The employed cavity mode $\rm{TE_{102}}$ has frequency $\omega_{\rm c}/2\pi=6.364~\rm{GHz}$. The static magnetic field applied to the YIG sphere induces strong magnetic-field fluctuations~\cite{Xu-PRL-2023}.
Consequently, we employ the Autler-Townes (AT) effect to dynamically tune the qubit frequency~\cite{Xu-PRL-2023,supplemental,AT-PR-1955} instead of achieving tunability by replacing the Josephson junction with a flux-sensitive SQUID~\cite{Fagaly-RSI-2003}. A strong microwave drive resonant with the $\vert e\rangle \rightarrow \vert f\rangle$ transition hybridizes the qubit levels into dressed states $\vert\pm\rangle=(\vert e\rangle\pm\vert f\rangle)/\sqrt{2}$. Therefore, the original transition from $\vert g \rangle$ to $\vert e \rangle$ splits into two branches (Fig.~1b). We define $\vert g\rangle$ and $\vert +\rangle$ as the two basis states of the dressed qubit, so the transition frequency of the qubit is $\omega_{\rm q} \approx \omega_{ge} + \frac{1}{2}\Omega_{\rm AT}$, where $\Omega_{\rm AT}$ is the Rabi frequency of the AT drive~\cite{Xu-PRL-2023,supplemental}. This AT frequency tuning allows for precise control of the qubit-magnon detuning $\Delta_{\rm qm} = \omega_{\rm m} - \omega_{\rm q}$, with $\omega_{\rm q}<\omega_{\rm m}$ here, so as to ensure both $\omega_{\rm q}$ and $\omega_{\rm m}$ largely detuned from the cavity mode $\rm{TE_{102}}$ (Fig.~1c). Under this condition, the cavity mode can be adiabatically eliminated, yielding an effective coupling $g_{\rm qm}$ between the qubit and the Kittel mode~\cite{supplemental}:
\begin{equation}\label{H0}
	H/\hbar=\frac{1}{2}\omega_{\rm q}\sigma_z+\omega_{\rm m}a^\dagger a+g_{\rm qm}(\sigma_+a+\sigma_-a^\dagger),
\end{equation}
where  $\sigma_+=\vert +\rangle\langle g\vert$, $\sigma_-=\vert g\rangle\langle +\vert$, and $a$ ($a^\dagger$) is the magnon annihilation (creation) operator.

To induce a strong magnon Kerr nonlinearity, we operate the system in a suitable magnon-qubit dispersive regime. Specifically, we set the detuning $\Delta_{\rm qm}\approx3.8g_{\rm qm}$, with $g_{\rm qm}/2\pi\approx 20.0~\rm{MHz}$, ensuring the qubit and magnon are far detuned but still interact strongly with each other via the exchange of virtual cavity photons. This interaction leads to an effective Kerr term in the magnon mode, which is essential for the quantum-level magnon squeezing. In this dispersive regime, we obtain an effective Hamiltonian~\cite{supplemental,Boissonneault-PRA-2009},
\begin{align} \label{Ht} 
	{H'}/{\hbar} &\approx (\omega_{\rm m} + \delta) a^\dagger a 
	+ \delta (a^\dagger a)^2 \sigma_z \notag \\
	&\quad + \biggl[\omega_{\rm q} - 2\chi \Bigl(a^\dagger a + \tfrac{1}{2}\Bigr)\biggr] \sigma_z/2,
\end{align}
where $\chi= g_{\rm qm}^2/\Delta_{\rm qm}-g_{\rm qm}^4/\Delta_{\rm qm}^3$ is the dispersive shift, and $\delta= g_{\rm qm}^4/\Delta_{\rm qm}^3$ is the self-Kerr coefficient. In the experiment, the qubit remains predominantly in $\vert g\rangle$ throughout the evolution, so we can project $\sigma_z$ onto the ground-state manifold ($\sigma_z=-1$) and reduce $H'$ to an effective Hamiltonian for the magnon
\begin{equation}\label{Heff}
	\begin{split}
		H_{\rm eff}/\hbar& = \omega'_{\rm m}a^{\dag}a - \delta(a^{\dag}a)^2,
	\end{split}
\end{equation}
where $\omega'_{\rm m}=\omega_{\rm m}+\chi+\delta \approx 2\pi \times 6.236$~GHz is the dressed magnon frequency. The crucial term, $-\delta(a^{\dag}a)^2$, describes the self-Kerr interaction, which is responsible for magnon squeezing. 

To understand the squeezing mechanism, we transform to a rotating frame at frequency $\omega'_{\rm m}$. In this frame, the Hamiltonian is converted to pure Kerr interaction: $H_{\rm Kerr}/\hbar=-\delta(a^{\dagger}a)^2$ \cite{supplemental}. The corresponding time-evolution operator, $U(t)=\exp[i\delta (a^\dagger a)^2t]$, imparts a phase shift proportional to $n^2$  to each magnon Fock state component $\vert n\rangle$. Consequently, higher-number Fock state components accumulate phase more rapidly, causing the magnon state to shear and distort in phase space.  This shearing process reshapes the state's phase-space distribution into a quasi-elliptical profile, which is characteristic of squeezing. Also, a microwave drive is required to pump the magnons to a non-vaccum state, as the vacuum state $\vert 0\rangle$ is an eigenstate of $H_{\rm Kerr}$ and remains unaffected by $H_{\rm Kerr}$. Notably, this Kerr-induced squeezing produces a distinctive curved profile, differing from the perfect ellipse generated by a standard squeezing operator $S(r)$ \cite{supplemental}.

The control sequence for magnon squeezing is given in Fig.~1d. The qubit-magnon system is initialized in the joint ground state $\vert g,0\rangle=\vert g\rangle\otimes\vert 0\rangle$, where $\vert0\rangle$ denotes the magnon vacuum state. To generate Kerr evolution, we apply a strong AT drive on the transmon, with an amplitude of $\Omega_{\rm AT}/2\pi = 76~\rm{MHz}$ and duration time $\tau$, which tunes the qubit frequency to $\omega_{\rm q}/2\pi = 6.155~\rm{GHz}$ during $\tau$. Therefore, the detuning becomes $\Delta_{\rm qm}/2\pi = 76~\rm{MHz}\approx$ $3.8g_{\rm qm}$, leading to a significant self-Kerr coefficient $\delta \approx 2\pi \times 0.36~\rm{MHz}$. Simultaneously, a microwave pulse resonant with the dressed magnon frequency and having an amplitude of $11.7~\rm{MHz}$ is also applied for a duration $\tau$, initiating the Kerr evolution to generate squeezed states $\rho_{\rm m}$ for the magnon mode.

\subsection{Wigner tomography and density matrix construction}

We perform Wigner tomography to verify the deterministic generation of squeezed magnon states. The magnon state $\rho_{\rm m}$ is displaced and then its parity measured: $W(\alpha)=(2/\pi) {\rm Tr}[D(-\alpha)\rho_{\rm m} D(\alpha)P]$, where $D(\alpha)=\exp(\alpha a^\dagger-\alpha^*a)$ is the magnonic displacement operator, and $P=\exp(i\pi a^\dagger a)$ is the magnonic parity operator. The displacement $D(\alpha)$ is implemented via a microwave pulse resonant with the magnon frequency, with $\alpha$ controlled by both the pulse amplitude and phase. 

Measuring the expectation value of the parity operator $P$ typically requires magnon-qubit state swapping. However, directly tuning the AT drive to achieve resonant magnon–qubit swapping is impeded by a substantial detuning ($\Delta_{ge,{\rm m}}/2\pi=(\omega_{\rm m}-\omega_{ge})/2\pi \approx 124~\rm{MHz}$) that is comparable to the qubit anharmonicity ($\eta/2\pi\approx-249~\rm{MHz}$). To circumvent this difficulty, we implement a novel magnon-assisted Raman process~\cite{Pechal-PRX-2014, supplemental} by applying an anxiliary pulse (with duration $t$) on the transmon at $\omega_d=2\omega_{ge}+\eta-\omega_{\rm m}$. This drive induces a second-order interaction that couples $\vert f,n\rangle$ and $\vert g,n+1\rangle$ with an interaction strength $g_n=g_0\sqrt{n+1}$, where $g_0$ is proportional to the drive amplitude $\Omega$~\cite{supplemental}, thus continuously tunable by $\Omega$~(Fig.~1e). In the rotating frame of the drive, $\vert f,n\rangle$ and $\vert g,n+1\rangle$ become resonant, enabling resonant magnon-qubit swapping through direct energy exchange~\cite{supplemental}. During this process, $\vert g\rangle$ and $\vert f\rangle$ act as an effective two-level system. The control sequence for implementing this resonant swapping is illustrated in Fig.~1d.

The results of the Wigner tomography are presented in Fig.~2a. The top row shows the experimentally measured Wigner functions for the magnonic vacuum state ($\tau = 0$) and squeezed states at various evolution times ($\tau = 50$, 100, 150, 200, and 250~ns). The bottom row displays the corresponding simulated results. From the measured Wigner functions in the top row, we obtain the density matrices of the magnon states, cf. dark blue bars in Fig.~2b, where only magnon states at $\tau=0$, 100, and 200~ns are presented for clarity. Numerically simulated density matrices (red frames), obtained via a Master equation approach~\cite{supplemental}, are also shown in Fig.~2b. From the numerical density matrices, we then reconstruct the Wigner functions in the bottom row of Fig.~2a. It is clear that the numerical simulations agree well with the experimental results, confirming the deterministic generation of magnon squeezing.

\subsection{Magnon squeezing characterization}

To further characterize the magnon squeezing, we evaluate the variance $V(\tilde{X})$ of a generic quadrature operator $\tilde{X}$ defined by
$\tilde{X}(\theta) = \cos(\theta) X + \sin(\theta) P$, where $\theta$ gives the angle of orientation in the phase space; $X = a + a^\dagger$, and $P = -i(a - a^\dagger)$ are the nominal position and momentum quadratures of the magnon, respectively. For $\theta$ from 0 to 2$\pi$, the corresponding variances are extracted from the density matrices (Fig.~2b). For instance, in Fig.~3a, we show the measured and simulated quadrature variances versus the rotation angle $\theta$ for the evolution times $\tau=$ 150 and 250~ns, respectively. For each evolution time $\tau$, we determine the minimum variance $V_{\rm min}$ at a given angle $\theta_{\rm min}$ and the corresponding maximum variance $V_{\rm max}$ at the angle $\theta_{\rm max}=\theta_{\rm min}+\pi/2$. Figure~3b presents the measured quadrature variances $V_{\rm min}$ (red circles) and $V_{\rm max}$ (orange squares) versus evolution time $\tau$. The smallest value of the measured minimum variances is $V_{\rm min}\approx 0.799 \pm 0.068$ at $\tau=150$~ns, corresponding to $\sim\!1.0$~dB squeezing, in good agreement with the numerical simulations (red curve)~\cite{supplemental}. Notably, even when accounting for one standard deviation (1 s.d.) errors, the measured minimum variances remain below the vacuum level ($V=1$).

To verify that the observed magnon squeezing indeed occurs in the quantum regime, we calculate the mean magnon number $\langle n\rangle=\sum_n n \rho_{\rm m}^{nn}$. The results are shown in Fig.~3c, demonstrating that the mean magnon number remains below 1 throughout the given evolution period. This confirms that the magnon squeezing is achieved at the quantum level of extremely low excitations, providing an unequivocal experimental evidence for the  quantum-level magnon squeezing in a macroscopic spin system. Note that for the magnon frequency $\omega_{\rm m}/2\pi=6.231$~GHz and the experimental temperature of $T < 20$~mK, the mean number of thermally excited magnons is $\langle n\rangle_{\rm th} < [\exp(\hbar\omega_{\rm m}/k_BT)-1]^{-1}\approx 3.2\times 10^{-7}$. This exceedingly small $\langle n\rangle_{\rm th}$ means that the influence of thermal excitation is negligible in our experiment: it is the quantum fluctuations of the magnons that are being manipulated and squeezed.

\begin{figure}
	\includegraphics[width=0.46	\textwidth]{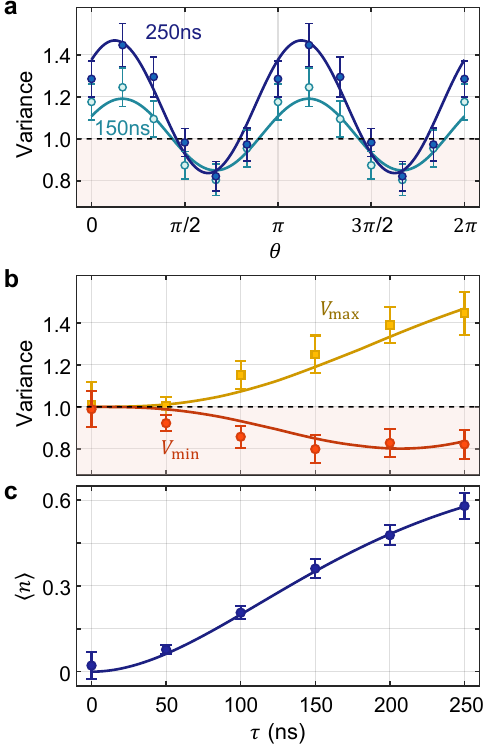}%
	\caption{\textbf{Quadrature variance and mean magnon number of the magnon state.} \textbf{a,} Quadrature variances $\tilde{X}(\theta)$ versus rotation angle $\theta$ for the evolution times $\tau=$ 150 and 250~ns. The light and dark blue circles (with 1 s.d. error bars) represent the variances that are extracted from the density matrices obtained from measured Wigner functions for $\tau=$ 150 and 250~ns, respectively. The corresponding light and dark blue solid curves show the numerical simulations. Dashed line is the ground state variance and the shaded region highlights quantum squeezing. \textbf{b,} Minimum and maximum quadrature variances versus evolution time $\tau$. Red circles (with 1 s.d. error bars) denote the measured minimum variance of the quadrature $\tilde{X}(\theta)$, and orange squares (with 1 s.d. error bars) represent the variance of its conjugate counterpart (the maximum quadrature variance). Solid curves show numerical simulations. Dashed line is the ground state variance and the shaded region highlights quantum squeezing. The largest squeezing of $\sim\!1.0$ dB appears at around 150~ns. Notably, the upper bounds of the error bars for the minimum variance (red circles) remain below 1.0. \textbf{c,} Mean magnon number $\langle n\rangle$ as a function of $\tau$. Filled circles (with 1 s.d. error bars) are experimental data, and the solid curve is the numerical simulation. The mean magnon number remains below one, a hallmark of extremely low excitations that is characteristic of magnon squeezing in the quantum regime.}
	\label{fig3}
\end{figure}

\subsection{Dynamics of the squeezed magnon state under decoherence}
Having prepared and characterized the squeezed magnon state, we investigate its time evolution under decoherence --- a critical step towards leveraging squeezed magnons in quantum technologies. 
	
We first characterize the intrinsic decay of the squeezed state. After the state generation, the AT drive is removed, which effectively eliminates the Kerr nonlinearity, $\delta \propto \Delta_{\rm qm}^{-3} \approx 0$, by maximizing the qubit-magnon detuning. The state is then allowed to evolve freely for a waiting time $\tau_{\rm w}$ before Wigner tomography is performed. The resulting Wigner functions, shown in the top row of Fig.~4a, reveal a gradual transition from a quasi-elliptical distribution --- the hallmark of squeezing --- to a circular profile of the vacuum state as $\tau_{\rm w}$ increases from 0 to 400 ns. A quantitative analysis (Fig.~4b, left column) confirms that both the minimum and maximum quadrature variances, $V_{\rm min}$ and $V_{\rm max}$, relax exponentially to the vacuum level of unity. The decay rate $\gamma_{\rm m}/2\pi=1.10~\rm{MHz}$ matches that of the mean magnon number $\langle n\rangle$, defining an intrinsic squeezed-state lifetime of $T_{1,{\rm m}}=1/\gamma_{\rm m}\approx145$~ns.

If the self-Kerr nonlinearity is maintained during the evolution, we can prolong the time interval over which squeezing is observable. In this protocol, the AT drive remains active after state preparation and the self-Kerr nonlinearity is tuned to $\delta/2\pi=0.25~$MHz~\cite{supplemental}. This persistent nonlinearity alters the decoherence pathway. While the mean magnon number still decays exponentially at the intrinsic rate $\gamma_{\rm m}$, the evolution of the quadrature variances deviates significantly from a simple exponential decay (Fig.~4b, right column). The nonlinearity actively counteracts dissipation, considerably slowing down the relaxation of variances to the vacuum level. As a result, at $\tau_{\rm w}=400$~ns, we still observe $V_{\rm min}(\tilde{X})\approx 0.957 \pm 0.029$. This measurement confirms that quantum squeezing is maintained for a time significantly longer than the intrinsic magnon lifetime $T_{\rm 1,m}$.
	
This prolonged observation of the magnon squeezing can be understood in terms of a competition between two processes. On the one hand, magnon dissipation drives the system towards the vacuum state. On the other hand, competing with this, the sustained self-Kerr evolution continuously re-establishes the squeezed phase-space distribution. Such an interplay between dissipation and nonlinear evolution makes the quadrature variances relax to unity at a much slower rate than the magnon population. Note that the duration for which squeezing can be maintained in this manner is ultimately limited by the intrinsic magnon linewidth $\gamma_{\rm m}$. It could be extended further in the future by improving the material quality and reducing the surface roughness of the YIG sphere~\cite{Tabuchi-CRP-2016,Spencer-PRL-1959}.

\begin{figure*}
	\includegraphics[width=0.8\textwidth]{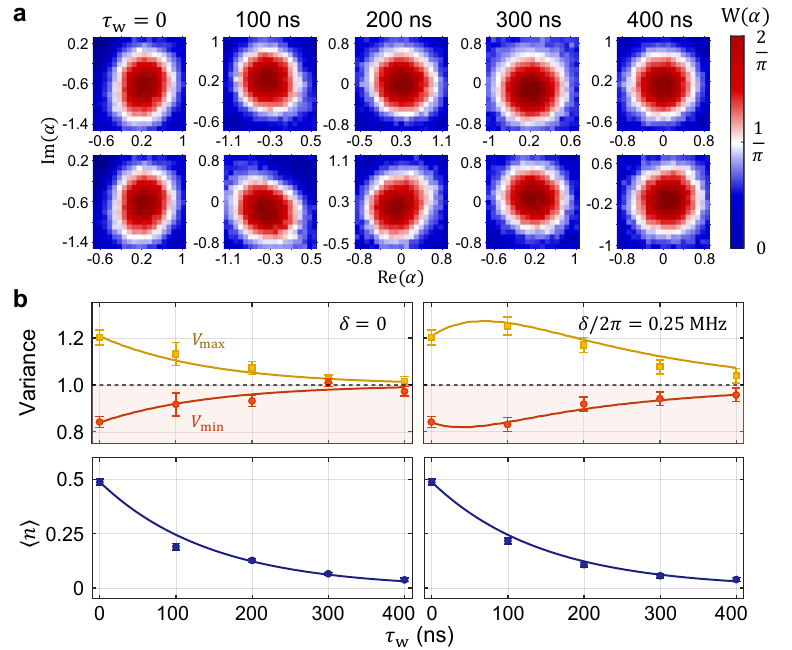}%
	\caption{\textbf{Decay and preservation of the squeezed magnon state.} 
		\textbf{a,}~Experimentally measured Wigner functions of the squeezed state evolving at selected waiting times $\tau_{\rm w}$. Top row: In the absence of nonlinearity ($\delta=0$), the initial squeezed state (elliptical distribution) gradually decays to the vacuum state $\vert 0 \rangle$ (circular distribution). Bottom row: With a persistent nonlinearity ($\delta/2\pi=0.25$~MHz), the phase-space distribution remains distinctly non-circular for a much longer duration, visually confirming the prolonged squeezing quantified below.
		\textbf{b,}~Squeezed state evolution. The left column shows the intrinsic decay when the Kerr nonlinearity is turned off ($\delta=0$), while the right column shows the evolution with a persistent nonlinearity ($\delta/2\pi=0.25$~MHz). Top row: Time evolution of the minimum ($V_{\rm min}$, red circles) and maximum ($V_{\rm max}$, orange squares) quadrature variances. Bottom row: Corresponding decay of the mean magnon number $\langle n\rangle$. Experimental data (points) are shown with 1 s.d. error bars. Dashed lines in the top panels indicate the ground state variance and the shaded regions highlight quantum squeezing. Solid curves in the top panels are numerical simulations, while those in the bottom panels are exponential fits, which consistently give rise to a magnon lifetime of $T_{1,m}\approx~$145~ns.}
	\label{fig4}
\end{figure*}

\section{Discussion}

In this study, we have demonstrated quantum-level magnon squeezing in a macroscopic millimeter-scale YIG sphere containing approximately $10^{19}$ spins, achieving up to 1.0 dB of magnon squeezing. Prior to our work, quantum squeezing has been realized in a variety of material systems, such as atomic ensembles \cite{Han-Nature-2020,Jacob-PRL-2023} and mechanical oscillators \cite{Marti-NP-2024,Wollman-Science-2015,Youssefi-NP-2023}. Also, there was evidence towards experimental observation of the parametrically squeezed states of microwave-frequency magnons in the YIG films~\cite{Kostylev-PRB-2019}, where the intrinsic nonlinearity of magnons in the YIG materials was used to generate the squeezed magnon states. As shown in \cite{Zhang-SCPMA-2019}, this intrinsic nonlinearity is too small to produce observable squeezing effects in the quantum regime. In contrast, our present experiment highlights the dispersive magnon-qubit coupling as a viable method to produce strong self-Kerr nonlinearity of magnons and to generate squeezed magnon states in the quantum regime. Indeed, as demonstrated in this experiment, the mean magnon number of the generated squeezed magnon states can be less than one. Furthermore, we demonstrate that the magnon self-Kerr nonlinearity can be harnessed to actively preserving quantum squeezing for a duration far exceeding the intrinsic magnon number decay time. Here, the observed magnon squeezing is primarily limited by the finite coupling strength $g_{\rm qm}$ and the intrinsic magnon linewidth $\gamma_{\rm m}$. The coupling strength could be enhanced by redesigning the cavity geometry or using YIG samples with a larger volume~\cite{Lachance-APE-2019}, while the intrinsic magnon linewidth could be extended with higher-purity YIG materials and more refined surface polishing~\cite{Tabuchi-CRP-2016}. Future improvements in these two aspects could therefore raise the quantum squeezing of magnons to a higher level.

Our work demonstrates the possibility of extending quantum-enhanced metrology to large-scale spin ensembles and testing the boundary between quantum and classical realms. This system could also serve as a versatile testbed to investigate quantum gravity effects with unprecedented mass and spin numbers~\cite{Bose-RMP-2025}. 
Moreover, the squeezing mechanism -- specifically, quantum squeezing induced by Kerr nonlinearity -- along with the techniques developed in this work (e.g., the magnon-assisted Raman process) may have broad applications not only in magnonic systems, but also in other hybrid platforms, such as the phonon-based systems~\cite{Marti-NP-2024,Hu-PRL-1996,Hu-PRL-1997}. The squeezed magnon states, in particular in the quantum regime, represent a possible resource for dark-matter axion searches~\cite{Crescini-PRL-2020,Mitridate-PRD-2020} and other ultrasensitive quantum metrology applications~\cite{Abadie-Np-2011}, revealing potentially tantalizing roles for macroscopic spin systems in next-generation quantum technologies.

\section{Methods}
\subsection{Device design}
As illustrated in Fig.~1a, the device consists of three parts: a single-junction 3D transmon, a 1 mm-diameter highly polished single-crystal YIG sphere, and a rectangular 3D microwave cavity. The transmon and the YIG sphere are placed near the electric- and magnetic-field antinodes of the $\rm{TE}_{102}$ mode, respectively, to maximize the coupling strength. The 3D cavity, with internal dimensions of $58\times41\times2~\rm{mm}^3$, is composed of two parts made of 6061 aluminum alloy and oxygen-free copper, respectively. At cryogenic temperatures, the superconductivity of the aluminum alloy reduces the cavity linewidth ($\kappa_{\rm{TE}102}/2\pi=0.73$~MHz) and protects the transmon from stray magnetic fields via the Meissner effect. Meanwhile, the oxygen-free copper part allows the bias magnetic field $\mathbf{B}$, which is perpendicular to the magnetic-field component of the $\rm{TE}_{102}$ mode, to penetrate this part of the cavity, saturating the magnetization of the YIG sphere. The bias magnetic field is composed of two components: a static part supplied by two neodymium iron-boron (NdFeB) square permanent magnetic plates (each of size $10\times10\times1~\rm{mm}^3$), and a tunable part provided by an electromagnet consisting of a niobium-titanium (NbTi) alloy superconducting coil with approximately 15000 turns wound around a pure iron core. The aluminum part of the 3D cavity is further covered by a 10 mm thick magnetic shield made of pure iron to protect the cavity.

\subsection{Measurement setup}
The whole experimental device is anchored to the base plate of a dilution refrigerator at $\sim10$~mK. Microwave signals used for qubit-magnon control are generated by digital-to-analog converter (DAC) boards at room temperature. They are then attenuated by a total of 53~dB, passed through an infrared filter and a 12.5~GHz low-pass filter on the MC plate for noise reduction, before entering the input port of the 3D cavity. The transmission signal from the output port of the 3D cavity is then passed through a 9.6~GHz low-pass filter and a dual isolator, amplified by a high-electron-mobility transistor (HEMT) cryogenic amplifier operating in the 4-8~GHz band on a 4K plate, and finally demodulated at room temperature by an IQ mixer before being fed into an analog-to-digital converter (ADC) board for analysis.

\section{Data Availability}
Source data for the figures in this paper have been deposited on Zenodo (\url{https://zenodo.org/records/18297252}). Additional information is available from the corresponding authors upon request.

\section{Acknowledgments}
This work is supported by the National Key Research and Development Program of China (Grant No.~2022YFA1405200 to J.Q.Y. and No.~2025YFE0201100 to D.X.), the National Natural Science Foundation of China (Grant No.~92265202 to J.Q.Y., No.~U25A20199 to J.Q.Y. and No.~12404563 to Z.C.), the ``Pioneer" and ``Leading Goose" R\&D Program of Zhejiang (Grant No. 2025C01028 to J.Q.Y.) and Beijing Natural Science Foundation (Grant No.~1244065 to Z.C.). H.F.Y. is supported by the National Natural Science Foundation of China (Grant No.~92365206) and The Innovation Program for Quantum Science and Technology (Grant No.~2021ZD0301802). X.H. acknowledges support by the CAS Dean's office at UB. F.N. is supported in part by the Japan Science and Technology Agency (JST) [via the CREST Quantum Frontiers program Grant No.~JPMJCR24I2, the Quantum Leap Flagship Program (Q-LEAP), and the Moonshot R\&D Grant No.~JPMJMS2061].

\section{Author contributions}
J.Q.Y. and D.X. conceived the experiment. Y.C.W. and D.X. designed and performed the experiment, acquired the data and carried out the data analysis under the supervision of J.Q.Y.. D.X. designed the cavity and qubit chip, Z.C. and H.Y. fabricated the qubit chip, J.L., X.H., S.Y.Z. and F.N. provided theoretical support, L.Z.T and X.K.G. provided technical support. All authors contributed to the discussion of the results and the writing of the manuscript.

\section{Competing Interests}
The authors declare no competing interests.

\section{Additional Information}
Correspondence and requests for materials should be addressed to Da Xu, Hai-Feng Yu or J. Q. You.


\section{References}

\begin{thebibliography}{999}
	
\bibitem{Walls-Nature-1983}
D. F. Walls, Squeezed States of Light, Nature {\bf 306}, 141-146 (1983).

\bibitem{Abadie-Np-2011}
J. Abadie et al., A Gravitational Wave Observatory Operating beyond the Quantum Shot-Noise Limit, Nature Phys {\bf 7}, 962-965 (2011).

\bibitem{Backes-Nature-2021}
K. M. Backes et al., A Quantum Enhanced Search for Dark Matter Axions, Nature {\bf 590}, 238-242 (2021).

\bibitem{Braunstein-RMP-2005}
S. L. Braunstein and P. van Loock, Quantum Information with Continuous Variables, Rev. Mod. Phys. {\bf 77}, 513 (2005).

\bibitem{Weedbrook-RMP-2012}
C. Weedbrook, S. Pirandola, R. García-Patrón, N. J. Cerf, T. C. Ralph, J. H. Shapiro, and S. Lloyd, Gaussian Quantum Information, Rev. Mod. Phys. {\bf 84}, 621 (2012).

\bibitem{Bose-RMP-2025}
S. Bose, I. Fuentes, A. A. Geraci, S. M. Khan, S. Qvarfort, M. Rademacher, M. Rashid, M. Toro\v{s}, H. Ulbricht, and C. C. Wanjura, Massive Quantum Systems as Interfaces of Quantum Mechanics and Gravity, Rev. Mod. Phys. {\bf 97}, 015003 (2025).

\bibitem{Flatte-PRL-2010}
{\"O}. O. Soykal and M. E. Flatt{\'e}, Strong Field Interactions Between a Nanomagnet and a
Photonic Cavity, Phys. Rev. Lett. {\bf 104}, 077202 (2010).

\bibitem{Huebl-PRL-2013}
H. Huebl, C. W. Zollitsch, J. Lotze, F. Hocke, M. Greifenstein, A. Marx, R. Gross, and S. T. B. Goennenwein,
High Cooperativity in Coupled Microwave Resonator Ferrimagnetic Insulator Hybrids, Phys. Rev. Lett. {\bf 111},
127003 (2013).

\bibitem{Nakamura-PRL-2014}
Y. Tabuchi, S. Ishino, T. Ishikawa, R. Yamazaki, K. Usami, and Y. Nakamura, Hybridizing Ferromagnetic Magnons and
Microwave Photons in the Quantum Limit, Phys. Rev. Lett. {\bf 113}, 083603 (2014).

\bibitem{Tang-PRL-2014}
X. Zhang, C.-L. Zou, L. Jiang, and H. X. Tang, Strongly Coupled Magnons and Cavity Microwave Photons, Phys.
Rev. Lett. {\bf 113}, 156401 (2014).

\bibitem{Tobar-PRApplied-2014}
M. Goryachev, W. G. Farr, D. L. Creedon, Y. Fan, M. Kostylev, and M. E. Tobar, High-Cooperativity Cavity
QED with Magnons at Microwave Frequencies, Phys. Rev. Applied {\bf 2}, 054002 (2014).

\bibitem{Hu-PRL-2015}
L. Bai, M. Harder, Y. P. Chen, X. Fan, J. Q. Xiao, and C.-M. Hu, Spin Pumping in Electrodynamically Coupled Magnon-Photon Systems, Phys. Rev. Lett. {\bf 114}, 227201 (2015).

\bibitem{Bauer-PRB-2015}
Y. Cao, P. Yan, H. Huebl, S. T. B. Goennenwein, and G. E. W. Bauer, Exchange magnon-polaritons in microwave
cavities, Phys. Rev. B {\bf 91}, 094423 (2015).

\bibitem{Zhang-QI-2015}
D. Zhang, X.-M. Wang, T.-F. Li, X.-Q. Luo, W. Wu, F. Nori, and J. Q. You, Cavity quantum electrodynamics with ferromagnetic magnons in a small yttrium-iron-garnet sphere, npj Quant. Inf. {\bf 1}, 15014 (2015).

\bibitem{Lachance-APE-2019}
D. Lachance-Quirion, Y. Tabuchi, A. Gloppe, K. Usami, and Y. Nakamura, Hybrid quantum systems based on magnonics, Appl. Phys. Express {\bf 12}, 070101 (2019).

\bibitem{Yuan-PR-2022}
H. Y. Yuan, Y. Cao, A. Kamra, P. Yan, R. A. Duine, Quantum magnonics: when magnon spintronics meets quantum information science, Phys. Rep. {\bf 965}, 1 (2022).

\bibitem{Rameshti-PR-2022}
B. Z. Rameshti, S. V. Kusminskiy, J. A. Haigh, K. Usami, D. Lachance-Quirion, Y. Nakamura, C.-M. Hu, H. X. Tang,
G. E.W. Bauer, and Y. M. Blanter, Cavity magnonics, Phys. Rep. {\bf 979}, 1 (2022).

\bibitem{Tabuchi-Science-2015}
Y. Tabuchi, S. Ishino, A. Noguchi, T. Ishikawa, R. Yamazaki, K. Usami, and Y. Nakamura, Coherent coupling between a ferromagnetic magnon and a superconducting qubit, Science {\bf 349}, 405-408 (2015).

\bibitem{Lachance-Quirion-SA-2017}
D. Lachance-Quirion, Y. Tabuchi, S. Ishino, A. Noguchi, T. Ishikawa, R. Yamazaki, and Y. Nakamura, Resolving Quanta of Collective Spin Excitations in a Millimeter-Sized Ferromagnet, Sci. Adv. {\bf 3}, e1603150 (2017).


\bibitem{Lachance-quirion-Science-2020}
D. Lachance-Quirion, S. Wolski, Y. Tabuchi, S. Kono, K. Usami, and Y. Nakamura, Entanglement-based single-shot detection of a single magnon with a superconducting qubit, Science {\bf 367}, 425-428 (2020).

\bibitem{Rani-arxiv}
S. Rani, X. Cao, A. E. Baptista, A. Hoffmann, and W. Pfaff, High Dynamic-Range Quantum Sensing of Magnons and Their Dynamics Using a Superconducting Qubit, arXiv:2412.11859.

\bibitem{Xu-PRL-2023}
D. Xu, X.-K. Gu, H.-K. Li, Y.-C. Weng, Y.-P. Wang, J. Li, H. Wang, S.-Y. Zhu, and J. Q. You, Quantum Control of a Single Magnon in a Macroscopic Spin System, Phys. Rev. Lett. {\bf 130}, 193603 (2023).

\bibitem{Xu-QST-2024}
D. Xu, X.-K. Gu, Y.-C. Weng, H.-K. Li, Y.-P. Wang, S.-Y. Zhu, and J. Q. You, Macroscopic Bell state between a millimeter-sized spin system and a superconducting qubit, Quantum Sci. Technol. {\bf 9}, 035002 (2024).

\bibitem{Elyasi-PRB-2020}
M. Elyasi, Y. M. Blanter, and G. E. W. Bauer, Resources of Nonlinear Cavity Magnonics for Quantum Information, Phys. Rev. B {\bf 101}, 054402 (2020).

\bibitem{Kamra-APL-2020}
A. Kamra, W. Belzig, and A. Brataas, Magnon-Squeezing as a Niche of Quantum Magnonics, Appl. Phys. Lett. {\bf 117}, 090501 (2020).

\bibitem{Imamoglu-PRL-2009}
A. Imamo{\u g}lu, Cavity QED based on collective magnetic dipole coupling: spin ensembles as hybrid two-level systems, Phys. Rev. Lett. \textbf{102}, 083602 (2009).

\bibitem{Marti-NP-2024}
Stefano Marti, Uwe von Lüpke, Om Joshi, Yu Yang, Marius Bild, Andraz Omahen, Yiwen Chu, and Matteo Fadel, Quantum squeezing in a nonlinear mechanical oscillator, Nat. Phys. {\bf 20}, 1448-1453 (2024).

\bibitem{Pechal-PRX-2014}
M. Pechal, L. Huthmacher, C. Eichler, S. Zeytinoğlu, A. A. Abdumalikov, S. Berger, A. Wallraff, and S. Filipp, Microwave-Controlled Generation of Shaped Single Photons in Circuit Quantum Electrodynamics, Phys. Rev. X {\bf 4}, 041010 (2014).

\bibitem{supplemental}
See Supplemental Material.

\bibitem{Crescini-PRL-2020}
N. Crescini et al., Axion Search with a Quantum-Limited Ferromagnetic Haloscope, Phys. Rev. Lett. {\bf 124}, 171801 (2020).

\bibitem{Mitridate-PRD-2020}
A. Mitridate, T. Trickle, Z. Zhang, and K. M. Zurek, Detectability of Axion Dark Matter with Phonon Polaritons and Magnons, Phys. Rev. D {\bf 102}, 095005 (2020).

\bibitem{Zhang-SCPMA-2019}
G. Q. Zhang, Y. P. Wang, and J. Q. You, Theory of the magnon Kerr effect in cavity magnonics, Sci. China-Phys. Mech. Astron. {\bf 62}, 987511 (2019).

\bibitem{Tabuchi-CRP-2016}
Y. Tabuchi, S. Ishino, A. Noguchi, T. Ishikawa, R. Yamazaki, K. Usami and Y. Nakamura, Quantum magnonics: The magnon meets the superconducting qubit, C. R. Phys. \textbf{17}, 729-739 (2016).

\bibitem{AT-PR-1955}
S. Autler and C. Townes, Stark effect in rapidly varying fields, Phys. Rev. \textbf{100}, 703 (1955).

\bibitem{Fagaly-RSI-2003}
R.L. Fagaly, Superconducting quantum interference device instruments and applications, Rev. Sci. Instrum. {\bf 77}, 101101 (2006).

\bibitem{Boissonneault-PRA-2009}
M. Boissonneault, J. M. Gambetta, and A. Blais, Dispersive Regime of Circuit QED: Photon-Dependent Qubit Dephasing and Relaxation Rates, Phys. Rev. A {\bf 79}, 013819 (2009).

\bibitem{Spencer-PRL-1959}
E. G. Spencer, R. C. LeCraw, and A. M. Clogston, Low-temperature line-width maximum in yttrium iron garnet, Phys. Rev. Lett. \textbf{3}, 32 (1959).

\bibitem{Han-Nature-2020}
H. Bao et al., Spin squeezing of $10^{11}$ atoms by prediction and retrodiction measurements, Nature \textbf{581}, 159-163 (2020).

\bibitem{Jacob-PRL-2023}
Jacob A. Hines et al., Spin squeezing by Rydberg dressing in an array of atomic ensembles, Phys. Rev. Lett. {\bf 131}, 063401 (2023).

\bibitem{Wollman-Science-2015}
E. E. Wollman, C. U. Lei, A. J. Weinstein, J. Suh, A. Kronwald, F. Marquardt, A. A. Clerk, and K. C. Schwab, Quantum squeezing of motion in a mechanical resonator, Science {\bf 349}, 952-955 (2015).

\bibitem{Youssefi-NP-2023}
Amir Youssefi, Shingo Kono, Mahdi Chegnizadeh, and Tobias J. Kippenberg, A squeezed mechanical oscillator with millisecond quantum decoherence, Nat. Phys. {\bf 19}, 1697-1702 (2023).

\bibitem{Kostylev-PRB-2019}
M. Kostylev, A. B. Ustinov, A. V. Drozdovskii, B. A. Kalinikos, and E. Ivanov, Towards experimental observation of parametrically squeezed states of microwave magnons in yttrium iron garnet films, 
Phys. Rev. B {\bf 100}, 020401(R) (2019).

\bibitem{Hu-PRL-1996}
X. Hu and F. Nori, Squeezed Phonon States:~Modulating Quantum Fluctuations of Atomic Displacements, Phys. Rev. Lett. {\bf 76}, 2294 (1996).

\bibitem{Hu-PRL-1997}
X. Hu and F. Nori, Phonon Squeezed States Generated by Second-Order Raman Scattering, Phys. Rev. Lett. {\bf 79}, 4605 (1997).

\end{thebibliography}
\end{document}